\documentclass[reqno,11pt]{amsart}
\usepackage{amssymb}

\newtheorem{lem}{Lemma}

\newtheorem{df}{Definition}
\newtheorem{hyp}{Hypothesis}

\newcommand\eps\varepsilon
\newcommand\ph\varphi
\newcommand\kap\varkappa

\begin{document}


\title[On Collisions in Nonholonomic systems]
{On Collisions in Nonholonomic systems}

\author[Dmitry Treschev and Oleg Zubelevich]{Dmitry Treschev and Oleg Zubelevich\\ \\\tt
 Dept. of Theoretical mechanics,  \\
Mechanics and Mathematics Faculty,\\
M. V. Lomonosov Moscow State University\\
Russia, 119899, Moscow, Vorob'evy gory, MGU \\
 }
\date{}
\thanks{Partially supported by grants
 RFBR 12-01-00441-à,  Science Sch.-2519.2012.1}

\begin{abstract}We consider  nonholonomic systems with  collisions and propose a concept of weak solutions to Lagrange-d'Alembert equations. In the light of this concept we describe dynamics of the collisions. Several applications have been investigated. Particularly the collision of rotating ball and the rough floor has been considered. 
 \end{abstract}

\maketitle
\numberwithin{equation}{section}
\newtheorem{theorem}{Theorem}[section]
\newtheorem{lemma}[theorem]{Lemma}
\newtheorem{definition}{Definition}[section]

\section{The Description of the Problem}

Let us start from the following model example. There is a solid ball $B$ of  radius $r$ and of mass $m$ and let its  centre of mass coincide with the geometric centre  $S$.  The moment of inertia relative to any axis passing trough the  point $S$ is equal to $J$.

Give an informal description of the problem.
 Being undergone with some potential forces the ball  rolls on the floor and sometimes it  collides with a vertical wall. After the collision it jumps aside the  wall. The wall and the floor are rough: the ball can not slide on the floor and along the wall.

We wish to construct a theory of such a motion in the  Lagrangian frame. Particularly, we wish to give sense to the term "elastic collision" in nonholonomic context.

In  physical space introduce a Cartesian coordinate system $Oxyz$. Let $(x_S,y_S,z_S)$ be the coordinates of the point $S$.

 Suppose that the plane $Oxy$ is a solid and rough floor and the plane $Ozy$ is a solid and rough wall. For all the time $t\ge 0$ we have $z_S=r, \quad x_S\ge r$.

By $C\in B$ denote the contact point of the ball and the floor. The ball can not slide on the floor:
\begin{equation}\label{sdrtvb5}\overline{v}_C=\overline v_S+[\overline \omega,\overline{SC}]=0,\end{equation} 
here $\overline v_C$ is the velocity of the point $C$, and $\overline \omega$ is the angle velocity of the ball.

When the ball reaches the wall (say by its point $G\in B$) then we also have
 \begin{equation}\label{sdrtvb5223}\overline{v}_G=\overline v_S+[\overline \omega,\overline{SG}]=0,\end{equation}

The configuration manifold of the system  is $M=\mathbb{R}^2\times SO(3)$, where $(x_S,y_S)\in \mathbb{R}^2$ and an element of $ SO(3)$ determines the orientation of the ball. We use the Euler angles for  the local coordinates in  $ SO(3)$ .

Consequently the position of the ball is determined by the vector $$x=(x_S,y_S,-\varphi,\theta,\psi)^T.$$ Why do we write $\varphi$ with negative sign will be clear below.  

The wall is a 4-dimensional manifold $N=\{x_S=r\}\subset M$.

Thus the general construction is as follows. We have a smooth configuration manifold $M,\quad \dim M=m$ and a smooth submanifold $N\subset M,\quad \dim N=m-1$ (the wall). Both manifolds carry the distributions. 

In the example under consideration the manifold $M$ carries the nonholonomic constraint given by  (\ref{sdrtvb5}) and the manifold $N$ carries the  constraint given by  (\ref{sdrtvb5}) and (\ref {sdrtvb5223}). 

Let $$x=( x_1,\ldots,x_m)^T\in M$$ be   local coordinates in $M$. 

To determine the distribution at each point $x\in M$ introduce a linear operator $$A(x):T_xM\to \mathbb{R}^{m-l},\quad\dim\mathrm{im}\,A(x)=m-l,\quad x\in M$$ and the mapping $x\mapsto A(x)$ is smooth. The subspaces $\ker A(x)\subseteq T_xM$ define an $l-$dimensional distribution in $M$.

To define an $s-$dimensional distribution in $N$ introduce a linear operator $$B(x):T_xM\to \mathbb{R}^{m-s},\quad\dim\mathrm{im}\,A(x)=m-s,\quad x\in N.$$
The distribution on $N$ consists of the subspaces $\ker B(x)\subseteq T_x N$. The operator $B(x)$ is also a smooth function of $x$.

The operators $A,B$ are not uniquely defined: the same distributions can be generated by the different operators $A,B$  but we use them because  they naturally arise in the applications.

Assume also that $$\ker B(x)\subseteq\ker A(x)$$ for each $x\in N$.

The dynamics of the system is described by the smooth Lagrangian $L(x,\dot x)$.

From the configuration manifold's geometry viewpoint the collisions of the rigid bodies  was  considered in \cite{KozlovTreschev}. 

One of the results of this article is as follows. The manifold $M$ is endowed with the Riemann metric generated by the kinetic energy of the system. The evolution of the system is expressed by the function $x(t)\in M$. When the point with coordinates $x(t)$ collides with the submanifold $N$ i.e. $x(\tau)\in N$ it then jumps aside obeying the law "the angle of incidence is equal to the angle of reflection".   The authors obtained this law from a system with non-solid constraint (wall) by means of some    natural limit process.

These results have been obtained in the absence of nonholonomic constraints. We generalize them  to the nonholonomic case.

In Section \ref{cvby6nu7} we consider a rough ball  colliding with a floor and obtain fromulas which particularly describe the following effect   \cite{Grawin}: "A perfectly rough ball which conserves kinetic energy behaves in such an unexpected way that it is difficult to pick up after it has bounced twice upon the floor, and, more bizarre, it returns to the hand on being thrown to the floor in such a way that it bounces from the underside of a table."

\section{The Weak Solutions to the Lagrange-d'Alembert Equation}

In the absence of unilateral  constraint $N$ a smooth function  $$x(t)=(x_1,\ldots,x_m)^T(t)\in M,\quad x_i(t)\in C^2 [t_1,t_2]$$ is the motion of the system if and only if for any function
\begin{align}\label{xdcfvrgtb}\psi(t)&=(\psi_1,\ldots,\psi_m)^T(t),\quad \psi_k\in \mathcal D(\mathbb{R}),\nonumber\\
\mathrm{supp}\,\psi_k&\subset(t_1,t_2),\quad \psi(t)\in \ker A(x(t))\end{align}
it satisfies the  Lagrange-d'Alembert equation
\begin{equation}\label{dcfvg5}\Big(\frac{\partial L}{\partial x}(x(t),\dot x(t))-\frac{d}{dt}\frac{\partial L}{\partial \dot x}(x(t),\dot x(t))\Big)\psi(t)=0,\quad t\in [t_1,t_2]\end{equation}
and the equation of constraint \begin{equation}\label{rcftg555}\dot x(t)\in \ker A(x(t)).\end{equation} 

\begin{df}We shall say that a function $x(t)\in H^1[t_1,t_2]$ is a weak solution to the system of Lagrange-d'Alembert equations and the equations of constraint iff the equation
\begin{equation}\label{c34rt5}\int_{t_1}^{t_2}\Big(\frac{\partial L}{\partial x}(x(t),\dot x(t))\psi(t)+\frac{\partial L}{\partial \dot x}(x(t),\dot x(t))\dot\psi(t)\Big)\,dt=0\end{equation}
holds for any $\psi$ that  satisfies (\ref{xdcfvrgtb}) and equation (\ref{rcftg555})  holds for almost all $t\in [t_1,t_2]$.
\end{df}

Note that by the Sobolev embedding theorem, the  space $H^1[t_0,t_1]$ belongs to $C[t_0,t_1]$.

In case of smooth function $x(t)$, equations (\ref{c34rt5}), (\ref{rcftg555}) are equivalent to equations (\ref{dcfvg5}), (\ref{rcftg555}). This follows from integration by parts and the Lagrange-d'Alembert principle \cite{bloch},\cite{Rum}.

 If the motion $x(t)$  contains collisions it is piece-wise differentiable: at the moment of collision its first derivative   is not continuous and the second one does not exist. 

Equations (\ref{c34rt5}) do not contain the second derivative of $x(t)$.
Therefore the concept of weak solutions is a proper tool to describe the motion with collisions.

 Let us turn to the details.

Consider  a solution $x(t)$ that collides the wall  at the moment $\tau\in (t_1,t_2)$ i.e. $x(\tau)\in N$. Correspondingly, one must put \begin{equation}\label{cfrtg56ff}\psi(\tau)\in \ker B(x(\tau)).\end{equation}

 We suppose that $x(t)\in C[t_1,t_2]$ and $$x(t)=\begin{cases}x^-(t),&t\in[t_1,\tau],\\
x^+(t),& t\in(\tau,t_2]\end{cases}$$ and $x^-(t)\in C^2[t_1,\tau],\quad x^+(t)\in C^2(\tau,t_2].$ 

By definition put $$x^+(\tau)=\lim_{t\to\tau+}x^+(t),\quad \dot x^+(\tau)=\lim_{t\to\tau+}\dot x^+(t).$$ We assume that these limits exist.

The solution $x(t)$ must obey nonholonomic constraint that is $$\dot x^\pm(t)\in\ker A(x^\pm(t)).$$ 
If $\dot x^-(\tau)=\dot x^+(\tau)$ then the derivative $\dot x(\tau)$ is defined and
\begin{equation}\label{qq}\dot x(\tau)\in \ker B(x(\tau)).\end{equation}

\subsection{The Equations of Collision}
Introduce the notation $$ v^{\pm}=\dot x^\pm(\tau)\in \ker A(x(\tau)).$$

Splitting the integral (\ref {c34rt5}) in sum $\int_{t_1}^\tau+\int_{\tau}^{t_2}$ and integrating them by parts obtain
\begin{align}
\int_{t_1}^{\tau}\Big(\frac{\partial L}{\partial x}(x(t),\dot x(t))-\frac{d}{dt}\frac{\partial L}{\partial\dot x}(x(t),\dot x(t))\Big)\psi(t)\,dt=0,\label{cvby6u}\\
\int_{\tau}^{t_2}\Big(\frac{\partial L}{\partial x}(x(t),\dot x(t))-\frac{d}{dt}\frac{\partial L}{\partial\dot x}(x(t),\dot x(t))\Big)\psi(t)\,dt=0,\label{cvby6uaa}\\
\Big(\frac{\partial L}{\partial\dot x}(x(\tau),v^+)-\frac{\partial L}{\partial\dot x}(x(\tau),v^-)\Big)\psi(\tau)=0.\label{cvbtn}\end{align}
Equations (\ref{cvby6u}), (\ref{cvby6uaa}) express that  the functions $x^\pm(t)$  satisfy to the  Lagrange-d'Alembert equations. By the assumption they also satisfy the equations of constraint: $\dot x^\pm(t) \in \ker A(x^\pm(t)) .$ 
That is before and after the collision the system obeys to the  Lagrange-d'Alembert equations and the equations of constraint.

Particularly, if the system is holonomic outside $N$  (i.e. $ A(x)=0$) then in their domains the functions $x^\pm(t)$  satisfy the Lagrange equations
$$\frac{\partial L}{\partial x}(x^\pm(t),\dot x^\pm(t))-\frac{d}{dt}\frac{\partial L}{\partial\dot x}(x^\pm(t),\dot x^\pm(t))=0.$$

Equation (\ref{cvbtn}) describes the behaviour of the system at the moment of collision. This equation is of main importance for us.

\section{Lemma from Vector Algebra}The following lemma is  mainly used in Section  \ref{ctvby6n}. But we placed it here because it provides an introduction to the geometry of the next Section. 
\begin{lem}\label{dcxfvgbt11}Let $L=\mathbb{R}^m$ be a Euclidean vector space with scalar product given by its Gramian matrix $G$. And let $B$ be the matrix of a linear operator (we denote operators and their matrices by the same letters) $$B:L\to\mathbb{R}^{m-s},\quad \mathrm{rang}\,B=m-s.$$
Let $$L=\ker B\oplus W,\quad W\perp \ker B$$ be the orthogonal decomposition of the space.

Then the square  matrix of orthogonal projector $P:L\to L,\quad P(L)=W$ is \begin{equation}\label{cvbhny7j}P=G^{-1}B^T\big(BG^{-1}B^T\big)^{-1}B.\end{equation}

If an operator $$A:L\to\mathbb{R}^{k}$$ is such that $\ker B\subseteq \ker A$ then one has \begin{equation}\label{cfvgbth}AP=A.\end{equation} Particularly, this implies that $P(\ker A)\subseteq \ker A$.\end{lem} 
{\it Proof.}  To obtain formula (\ref{cvbhny7j}) fix an arbitrary vector $x\in L$ and introduce a linear function $f(\xi)=(Px)^TG\xi.$ It is clear $$\ker B\subseteq\ker f.$$ This implies that there is an operator $\lambda:\mathbb{R}^{m-s}\to \mathbb{R}$ such that $(Px)^TG=\lambda B$ and $Px=G^{-1}B^T\lambda^T.$ It remains to find $\lambda^T$ from the equation $B(x-Px)=0$.

To obtain formula (\ref{cfvgbth}) note that there exists an operator $$\gamma:\mathbb{R}^{m-s}\to \mathbb{R}^{k}$$ such that $A=\gamma B$. Consequently, formula (\ref{cfvgbth}) follows from (\ref{cvbhny7j}).

The Lemma is proved.

\section{The Natural Lagrangian System}

To proceed with our analysis put
$$L=T(x,\dot x,\dot x)-V(x).$$ The form $$T(x,\xi,\eta)=\frac{1}{2}\xi^TG(x)\eta,\quad \xi=(\xi_1,\ldots,\xi_m)^T,\quad \eta=(\eta_1,\ldots,\eta_m)^T $$ is the kinetic energy of the system, the  matrix $ G(x)\equiv G^T(x)$ is positive definite. It defines  a Riemann metric in $M$. The potential energy $V$ is a smooth function in $M$.

By  (\ref{cfrtg56ff})   equation  (\ref{cvbtn}) is reduced to
\begin{equation}\label{dcvbb}T(x(\tau),v^+-v^-,u)=0,\quad u\in\ker B(x(\tau)).\end{equation}
From formula (\ref{dcvbb}) it follows that the difference $v^+-v^-$ is perpendicular to $\ker B(x(\tau))$.

\begin{hyp}\label{cfgvr6bhbv} The vector  $v^+$  depends on the vector $v^-$ by means of a linear operator $$v^+=R(x(\tau))v^-,\quad R(x(\tau)): \ker A(x(\tau))\to \ker A(x(\tau)).$$\end{hyp}

\begin{hyp}\label{fvg5bg}The energy is conserved during the collision:
$$T(x(\tau),v^+,v^+)=T(x(\tau),v^-,v^-).$$\end{hyp}

\begin{hyp}\label{cvb65}The system is reversible: if $x(t)$ is a motion of the system then $x(-t)$ is also a motion. For the collision this implies that:
$$v^-= R(x(\tau))v^+.$$\end{hyp}The third Hypothesis implies $(R(x(\tau)))^2=I$.

Note that if $\dim\ker B(x(\tau))=\dim\ker A(x(\tau))-1$ then the last Hypothesis is fulfilled automatically.

It is reasonable to consider  the following decomposition
$$T_{ x(\tau)}M=\ker B(x(\tau))\oplus W(x(\tau)),$$ here $W(x(\tau))$ is the orthogonal complement for $\ker B(x(\tau))$, and let $$P: T_{ x(\tau)}M\to W(x(\tau))$$ is the orthogonal projection. 

Introduce  notations $Pv=v_\perp,\quad (I-P)v=v_\parallel\in\ker B(x(\tau))$ and the norm  $|\xi|^2=T(x(\tau),\xi,\xi)$. Then write
$$v^{\pm}=v^{\pm}_\perp+v^{\pm}_\parallel.$$ 
\begin{theorem}\label{sdxcv5tb6y}Under hypotheses \ref{cfgvr6bhbv}-\ref{cvb65}
the following formula holds
\begin{equation}\label{fgvbh633333}v^+=(I-2P)v^-.\end{equation}
\end{theorem}
\begin{theorem}\label{dfergy}Formula (\ref{fgvbh633333}) gives physically correct model of collision.

Namely, denote by $x(t,\hat x,\hat v),\quad t\in[t_1,t_2]$ a  solution  such that 
$$ x(t_1,\hat x,\hat v)=\hat x,\quad \dot x(t_1,\hat x,\hat v)=\hat v$$ and for
some $\hat x',\hat v'$ and $\tau'\in(t_1,t_2)$ we have $$\lim_{t\to\tau'-}\dot x(t,\hat x',\hat v')\notin T_{x(\tau',\hat x',\hat v')}N,\quad x(\tau',\hat x',\hat v')\in N .$$ 
 
Then   the solution $x(t,\hat x,\hat v)$ is a continuous function of $t\in[t_1,t_2]$ and $(\hat x,\hat v)$ close to $(\hat x',\hat v')$, and there exists a continuous function $$\tau=\tau(\hat x,\hat v),\quad \tau(\hat x',\hat v')=\tau'$$ such that the collision occurs  at the moment $\tau$ i.e.  $x(\tau(\hat x,\hat v),\hat x,\hat v)\in N$.\end{theorem}
 Theorem \ref{dfergy} follows directly  from the Implicit Function Theorem and from the fact that the function $x\mapsto B(x)$ is continuous. 

\subsubsection{Proof of Theorem \ref{sdxcv5tb6y}}
Actually we deal  with vectors $v\in\ker A(x(\tau))$ only. By Lemma \ref{dcxfvgbt11} one has $P(\ker A(x(\tau))\subseteq \ker A(x(\tau))$.

 By Hypothesis \ref{fvg5bg} it follows that $|v^+|=|v^-|$ and  $R(x(\tau))$ is an isometric operator.

Since the difference $v^+-v^-=(v^+_\parallel-v^-_\parallel)+(v^+_\perp-v^-_\perp)$ is perpendicular to $\ker B(x(\tau))$ we have $$v^+_\parallel=v^-_\parallel$$ so that
$$R(x(\tau))\mid _{\ker B(x(\tau))}=I.$$
Introduce the following space $$F(x(\tau))=W(x(\tau))\cap\ker A(x(\tau))$$
then $\ker A(x(\tau))=F(x(\tau))\oplus\ker B(x(\tau))$ and $R(x(\tau))F(x(\tau))=F(x(\tau))$.

We finally have \begin{equation}\label{fgfgfggvgbt}v^+=Qv^-_\perp+v^-_\parallel,\quad Q=R(x(\tau))\mid_{F(x(\tau))}:F(x(\tau))\to F(x(\tau)).\end{equation}

Hypothesis \ref{cvb65}  implies that $Q^2=I$ and consequently each eigenvalue of $Q$ is  either equal to $1$ or to $-1$.

Show that the equality $Qv^-_\perp=v^-_\perp$ is possible if only $v^-_\perp=0$. Indeed, this equality implies $v^+=Qv^-_\perp+v^-_\parallel=v^-$. According to formula (\ref{qq}) one has $v^-\in \ker B(x(\tau))$ so that $v^-_\perp=0.$ Consequently $Q=-I$ and
$v^+=-v^-_\perp+v^-_\parallel.$
In terms of the matrix $P$ the same is written as (\ref{fgvbh633333}).

The Theorem is proved.

\section{Applications}\label{ctvby6n}Introduce the following notations $J'=J+r^2m,\quad \tilde J=J+r^2m/2.$
\subsection{The Ball Rolls on the Floor and Meets the Wall}In this section we solve the problem  we started with.

Let $\overline v_S^\pm,\quad \overline \omega^\pm$ stand for velocity of the point $S$ and for the ball's angular velocity respectively.  Superscripts $+$ and $-$ mark the states after the collision and before the collision respectively.

In the coordinates $Oxyz$ one has : $$\overline v_S^{\pm}=v_1^{\pm}\overline e_x+v_2^{\pm}\overline e_y,\quad\overline\omega^{\pm}=
\omega_1^{\pm}\overline e_x+\omega_2^{\pm}\overline e_y+\omega_3^{\pm}\overline e_z.$$

From formula (\ref{sdrtvb5}) one has 
\begin{equation}\label{cvbyt}v_1^\pm=\omega_2^\pm r ,\quad v_2^\pm=-\omega_1^\pm r.\end{equation}Therefore the velocity of any point of the ball is completely defined by quantities  $v_1^\pm,v_2^\pm,\omega_3^\pm.$
\begin{theorem}\label{vbhn76} The following formalas hold
\begin{align}
v_1^+=&-v_1^-,\nonumber\\
v_2^+=&\frac{r^2m}{2\tilde J}v_2^-+\frac{rJ}{\tilde J}\omega^-_3,\nonumber\\
\omega_3^+
=&\frac{J'}{r\tilde J}v_2^--\frac{r^2m}{2\tilde J}\omega_3^-.
\nonumber\end{align}\end{theorem}
{\it Proof.}
Introduce the Euler angles  so that at the moment of collision one has
 $\varphi=\psi=0,\quad \theta=\pi/2$. Then it follows that $\overline\omega=\dot\theta\overline e_x+\dot\psi\overline e_z-\dot\varphi\overline e_y.$
 Thus at the moment of collision we have $$v^{\pm}=(v^{\pm}_1,v^{\pm}_2,\omega^{\pm}_1,\omega^{\pm}_2,\omega^{\pm}_3)^T.$$
The formula $T=\frac{1}{2}m\overline v_S^2+\frac{1}{2}J\overline \omega^2$ implies $G=\mathrm{diag}(m,m,J,J,J)$.

Combining formulas (\ref{cvbyt}) and (\ref{sdrtvb5223}) we obtain
$$A=\begin{pmatrix}
  1  &0&0&-r&0 \\
  0 & 1 &r&0&0\\
 \end{pmatrix},\quad B=\begin{pmatrix}
  1  &0&0&0&0 \\
  0 & 1 &0&0&-r\\
0&0&0&1&0\\
0&1&r&0&0
 \end{pmatrix}.$$
The matrix $P$ is calculated with the help of Lemma \ref{dcxfvgbt11}:
$$P=\frac{1}{2\tilde J}\begin{pmatrix}
  1&0&0&0&0 \\
  0&2J&Jr&0&-Jr\\
0&rm&J'&0&J\\
0&0&0&1&0\\
0&-rm&J&0&J'
 \end{pmatrix}.$$

 Now the Theorem  follows  by direct calculation from  formula (\ref{fgvbh633333}).

The Theorem is proved.
\subsection{The  Ball is Thrown  to the Floor}\label{cvby6nu7}

In this section we consider another problem with the ball. Now we have only the floor $Oxy$ and there is no wall.

Being undergone with some potential forces the ball can move in the half-space $\{z_S> r\}$  and sometimes it can collide with the floor. 

After the ball meets the floor ($z_S=r$) it then jumps aside. The point of contact $C\in B$ has the zero  velocity (\ref{sdrtvb5}).

Introduce the configuration manifold of our system  as $M=\mathbb{R}^3\times SO(3)$, where $(x_S,y_S,z_S)\in \mathbb{R}^3$ and an element of $ SO(3)$ states the orientation of the ball. As a local coordinates in  $ SO(3)$ we use the Euler angles. 
The floor is a fifth dimensional submanifold $N\subset M$ which is given by the equation $z_S=r$.

Let $\overline v_S^+,\quad \overline \omega^+$ stand for velocity of the point $S$ and for ball's angular velocity after the collision respectively.  Let $\overline v_S^-,\quad \overline \omega^-$ stand for  velocity of the point $S$ and for  ball's angular velocity before the collision respectively.

In the coordinates $Oxyz$ one has : $$\overline v_S^{\pm}=(v_1^{\pm},v_2^{\pm},v_3^{\pm}),\quad\overline\omega^{\pm}=
(\omega_1^{\pm},\omega_2^{\pm},\omega_3^{\pm}).$$

 \begin{theorem}\label{cvb6}
The following formulas hold true
\begin{align}
v_1^+&=\frac{mr^2- J}{J'}v_1^-+\frac{2Jr}{J'}\omega_2^-,\nonumber\\
v_2^+&=\frac{mr^2- J}{J'}v_2^--\frac{2Jr}{J'}\omega_1^-,\nonumber\\
v_3^+&=- v_3^-,\nonumber\\
\omega_1^+&=-\frac{2rm}{J'}v_2^-+
\frac{J- mr^2}{J'}\omega_1^-,\nonumber\\
\omega_2^+&=\frac{2rm}{J'}v_1^-+\frac{J- mr^2}{J'}\omega_2^-,\nonumber\\
\omega_3^+&=\omega_3^-.\nonumber
\end{align}
\end{theorem}

Note that by these formulas the angular momentum  about the point of contact $C$ is conserved during the collision:$$m[\overline{CS},\overline v_S^+]+J\overline \omega^+=m[\overline{CS},\overline v_S^-]+J\overline \omega^-.$$ Due to formula (\ref{cvbtn}) this is not surprise.

{\it Proof of the Theorem.} Introduce the Euler angles  so that at the moment of collision one has
 $\varphi=\psi=0,\quad \theta=\pi/2$. Then it follows that $\overline\omega=\dot\theta\overline e_x+\dot\psi\overline e_z-\dot\varphi\overline e_y.$
 Thus at the moment of collision we have $$v^{\pm}=(v^{\pm}_1,v^{\pm}_2,v^{\pm}_3,\omega^{\pm}_1,\omega^{\pm}_2,\omega^{\pm}_3)^T.$$
The formula $T=\frac{1}{2}m\overline v_S^2+\frac{1}{2}J\overline \omega^2$ implies $G=\mathrm{diag}(m,m,m,J,J,J)$. From formula (\ref{sdrtvb5}) one obtains  
$$B=\begin{pmatrix}
  1 & 0 &0&0&-r&0 \\
  0 & 1 &0&r&0&0\\
 0 & 0 &1&0&0&0
 \end{pmatrix},\quad A=0.$$ The matrix of the operator $P$ is computed with the help of Lemma \ref{dcxfvgbt11}:
$$P=\frac{1}{J'}\begin{pmatrix}
  J & 0 &0&0&-Jr&0 \\
  0 & J &0&Jr&0&0\\
 0 & 0 &1&0&0&0\\
0&rm&0&r^2m&0&0\\
-rm&0&0&0&r^2m&0\\
0&0&0&0&0&0
 \end{pmatrix}.$$

Now  Theorem \ref{cvb6} follows from  formula (\ref{fgvbh633333}).

The Theorem is proved.

\subsubsection{Nonholonomic Pendulum}
Suppose that the ball  moves in the standard gravity field $\overline g=-g\overline e_z.$

Throw the ball to the floor so that $$\overline v^-_S=-v\overline e_x-u\overline e_z,\quad \overline\omega^-=\frac{rmv}{J}\overline e_y,\quad u,v>0.$$
From  Theorem \ref{cvb6} it follows that
$$\overline v_S^+=-\overline v_S^-,\quad \overline\omega^+=-\overline\omega^-.$$ Thus after the ball jumped up from the floor its centre $S$ moves along the same parabola  just in the opposite direction. Since the angle velocity also changes its direction  the periodic motion begins. The ball knocks the floor jumps up and go down along the parabola, knocks the floor at another point then flies along the same parabola to the initial point and so on.

\section{A Remark on Inelastic Collision}The developed above theory allows to construct different models of inelastic collision. 

For example, in the spirit of Newton's law of restitution, 
propose the following hypothesis:
\begin{equation}\label{fvgb6hn7}v^+=-\mu v^-_\perp+v^-_\parallel,\end{equation} here $\mu\in [0,1]$ is the restitution coefficient. This hypothesis is consistent with (\ref{dcvbb}).

For plastic collision $\mu=0$ and for super elastic one $\mu=1$.
In terms of operator $P$ formula (\ref{fvgb6hn7}) has the form
$$v^+=(I-(1+\mu)P)v^-.$$

Under this hypothesis  the corresponding formulas for the ball colliding the floor take the form
\begin{align}
v_1^+&=\frac{mr^2-\mu J}{J'}v_1^-+\frac{Jr(1+\mu)}{J'}\omega_2^-,\nonumber\\
v_2^+&=\frac{mr^2-\mu J}{J'}v_2^--\frac{Jr(1+\mu)}{J'}\omega_1^-,\nonumber\\
v_3^+&=-\mu v_3^-,\nonumber\\
\omega_1^+&=-\frac{rm(1+\mu)}{J'}v_2^-+
\frac{J-\mu mr^2}{J'}\omega_1^-,\nonumber\\
\omega_2^+&=\frac{rm(1+\mu)}{J'}v_1^-+\frac{J-\mu mr^2}{J'}\omega_2^-,\nonumber\\
\omega_3^+&=\omega_3^-.\nonumber
\end{align}


 \end{document}